\documentclass[12pt,a4paper]{article}

\usepackage{graphicx}

\graphicspath{{Fig/}}
\usepackage{amsmath}
\numberwithin{equation}{section}
\begin{document}

\title{Nucleon structure function in noncommutative space-time}
\author{A. Rafiei, Z. Rezaei, A. Mirjalili\\Physics department, Yazd university, 89195-741, Yazd, Iran}
\maketitle

\abstract{In the context of noncommutative space-time, we investigate the nucleon structure functions which plays
an important role to identify the internal structure of nucleons. We use the corrected vertices and employ new
vertices that appear in two approaches of noncommutativity and calculate the proton structure functions in terms
of noncommutative tensor $\theta_{\mu\nu}$. To check our result, we plot the nucleon structure function (NSF),
$F_2(x)$, and compare it with experimental data and the result coming out from the GRV, GJR and CT10
parametrization models. We show that new vertex which is arising the noncommutativity correction will lead us to
better consistency between theoretical result and experimental data for NSF. This consistency would be better at
small values of $x$-Bjorken variable. To indicate and confirm the validity of our calculations, we
also act conversely and obtain an lower bound for the numerical values of $\Lambda_{NC}$ scale which are corresponding to the recent  reports.}

\section{Introduction}
\label{sec:intro}

Lepton-nucleon deep inelastic scattering (DIS) is an important tool to investigate nucleons and their
constituents. Nucleon structure functions are the physical quantities for this purpose. Many phenomenological
models have been established to investigate the structure functions of nucleons~\cite{mirjalili1,mirjalili2,sf1,sf2,sf3,sf4,sf5,sf6} but there is, however, small deviation between experimental
data and models' predictions. On the other hand, it is possible to search to investigate new physics, such as
noncommutative (NC) space-time, in DIS processes.
\\
 The motivation to consider noncommutative field theory (NCFT) come back to string theory, where it was shown
 that, in the presence of an constant background field, the end points of an open string have noncommutative
 space-time (NCST) properties~\cite{seiberg1,seiberg2}. \\
 There is a wide range for the energy scale of NCST ($\Lambda_{NC}$). This range is arising out  different models
 while includes similar vertices. Different results for numerical values of $\Lambda_{NC}$ from different models
 with similar vertices are due to the different employed experiments in related analysing process. Theses
 experiments include low energy one  as well as precise high energy collider experiments
  and finally sidereral and  astrophysical events~\cite{TeV1}. More description of them, are as following:
 \begin{itemize}
   \item At low energy  experiments, for instance, lamb shift in hydrogen~\cite{Lamb}, magnetic moment of muon~\cite{MM1,MM2,MM3}, atomic clock measurements~\cite{ACM} and Lorentz violation by clock comparison
       test~\cite{TeV9} have already been studied  in the presence of NCST.
       In three body bound state, the experimental data for a helium atom put an upper bound on the magnitude
       of the parameter of noncommutativity, $\theta_{NC}$~\cite{Helium}.
   \item At High energy collider experiments we can refer for example to  forbidden decays in standard model
       (SM) such as $Z\rightarrow\gamma\gamma$~\cite{Buric}, top quark decays~\cite{topq1,topq2,topq3},
       compton scattering~\cite{mathews}  which  have been investigated in NCST. In the experiment which has
       been done by  OPAL collaboration, NC bound from $e^- e^+$ scattering at 95\% CL is $\Lambda_{NC}> 141
       GeV$~\cite{opal}.
   \item Astrophysics and cosmological bounds on NCST  have also been explored in various processes, such as
       Energy Loss via $\gamma \rightarrow \bar{\nu}\nu$ in stellar clusters~\cite{schupp}, effects of $\gamma
       \rightarrow \bar{\nu}\nu$ in primordial nucleosynthesis~\cite{Horvat} and ultra high energy
       astrophysical neutrinos~\cite{AHEN}.
 \end{itemize}
As we have mentioned and according to articles which  have been cited, the bound on $\Lambda_{NC}$ is strongly
model dependent and for collider scattering experiments it is about a few $TeV$. \\
Some of collider searches about NCST can be qualified, considering  some significant references. In ref.~\cite{TeV8} NC effects in several $2\rightarrow2$ processes in $e^- e^+$ collisions such as Moller and Bhabha
scattering, pair annihilation and $\gamma\gamma\rightarrow\gamma\gamma$ scattering are investigated. As a result,
the NC scale about $1 TeV$ is extracted at high energy linear colliders. The pair production of neutral
electroweak gauge boson is studied at the LHC~\cite{TeV6} and it is showen that under conservative assumptions,
NC bound is $\Lambda_{NC}\geq 1 TeV$. Also pair production of charged gauge boson at the LHC~\cite{ohl} exhibit clear deviation for the azimuthal distribution from SM at $\Lambda_{NC}= 700 GeV$. The NC effect for
Drell-Yan process at the LHC has been taken into account in~\cite{phenomenological12}  and consequently the
related scale is explored such as $\Lambda_{NC}\geq 0.4 TeV$.\\
Two approaches have been suggested to construct noncommutative standard model(NCSM)~\cite{cpst,sw}.  Using these
approaches, Feynman rules have been derived in~\cite{melic-electroweak,melic-strong,haghighat,haghighat-batabi}
which have been used to search for phenomenological aspects of NCSM~\cite{phenomenological1,phenomenological2,phenomenological3,phenomenological4,phenomenological5,phenomenological6,phenomenological7,phenomenological8,
 phenomenological9,phenomenological10,phenomenological11,TeV5,TeV8n}. The significant features of NCSM is that
 there are not only NC corrections for existing vertices, that people use them to calculate the DIS processes,
 but also it contains new gauge boson interactions, that may cause some corrections at leading order
 approximation of perturbative QCD. Here we would like to employ NC corrections and the new arised interactions
 to do some phenomenological tasks for electron-proton scattering.\\
The organization of this paper is as following. In section~\ref{section.intro}  we make a brief remarks about NCSM.
In section~\ref{section.eN} electron-proton DIS is computed in two approaches of NCSM. In section~\ref{section.result} we take into account the
amended
parton distributions, based on the  NCSM approaches, to extract the nucleon structure function,
using GRV, GJR and CT10 parametrization models~\cite{GRV,GJR,CT10}. Finally we will summarize our results and
give our conclusion in section~\ref{section.conclusion}.
\section{Noncommutative Standard model}\label{section.intro}

Noncommutative theory leads to commutation relation between space-time coordinate
\begin{equation}
[\,{\hat x^\mu }\,,\,{\hat x^\nu }] = i{\theta ^{\mu \nu }},
\label{equation.1}
\end{equation}
 where hatted quantities are hermitian operators and ${\theta ^{\mu \nu }}$ , is real, constant and asymmetry
 tensor. A simple way to construct NCFT is the Weyl-Moyal star product~\cite{modore-riad1,modore-riad2}
\begin{equation}
(f * g)(x) = {\left. {\exp (\frac{1}{2}i{\theta ^{\mu \nu }}\,\frac{\partial }{{\partial {x^\mu }}}\frac{\partial
}{{\partial {y^\nu }}})f(x)\,g(y)} \right|_{y \to x}}
\label{equation.2}
\end{equation}
Substituting star product with usual multiplication between conventional fields will lead to NCFT. Star
production has no effect on the integral of quadratic term, i.e $\int {{d^4}x f*g}=\int {{d^4}x fg}$, thus
propagators are equal in both NCSM and SM~\cite{modore-riad1,modore-riad2}. This mechanism makes some
difficulties such as charge quantization (that restrict charges of matter fields to $0$,$\pm1$~\cite{Hayakawa1,Hayakawa2}), and definition of gauge group tensor product~\cite{no go theorem}.

 Two approaches are suggested to resolve these problems. The first one is built from U(n) gauge group that is a
 bigger group with respect to the symmetry groups of standard model. On this base two Higgs mechanism reduces to
 standard model group~\cite{cpst}(we call this approach as unexpanded approach). The second one is based on
 Seiberg-Witten map~\cite{seiberg1,seiberg2} that gauge group is like the one of the standard model and
 non-commutative fields are expanded in terms of commutative ones (we call it expanded approach)~\cite{sw}. \\
 It is obvious that to consider a prefered direction makes the violation of Lorentz invariance. Also it had shown
 that noncommutative field theories are not unitary for ${\theta ^{\mu 0}} \ne 0$ , therefore, for observable
 measurements we should take ${\theta ^{\mu 0}} = 0$~\cite{mehen}.

 As previously mentioned, Feynman rules have derived in both approaches~\cite{melic-electroweak,melic-strong,haghighat,haghighat-batabi}. All vertices contain NC corrections. In
 addition, there are some new interactions. For example for electromagnetic interaction between lepton and
 proton, there are corrections in $l \gamma l$ and $q \gamma q$ as lepton and quark vertices.  In SM photon does
 not interact with neutral particles like neutrino, gluon and etc, but these interactions would exist in NCFT.
 Therefore one of new and outstanding vertices is photon-gluon interaction. Photon-fermion and photon-gluon
 vertices can be described briefly in different approaches as following.

\begin{itemize}

\item In expanded approach:

\begin{enumerate}

\item Photon-fermion vertex will be given by the expression as in below~\cite{melic-electroweak}:

\begin{equation}\label{eqnarray.photonfermion}
   \begin{array}{l}
i\,e\,{Q_f}\left[ {{\gamma _\mu } - \frac{i}{2}{q^\nu }\left( {({\theta _{\mu \nu }}{\gamma _\rho } + {\theta
_{\nu \rho }}{\gamma _\mu } + {\theta _{\rho \mu }}{\gamma _\nu })p_{in}^\rho  - {\theta _{\mu \nu }}{m_f}}
\right)} \right]\\
 = i\,e{Q_f}{\gamma _\mu }\\
\quad \; + \frac{1}{2}e{Q_f}\left[ {({p_{out}}.\theta .{p_{in}}){\gamma _\mu } - ({p_{out}}.{\theta _\mu
})({{\not p}_{in}} - {m_f}) - ({{\not p}_{out}} - {m_f})({\theta _\mu }.{p_{in}})} \right],
\end{array}
\end{equation}

\item Photon-gluon vertex is given by\cite{melic-strong}:
\begin{equation}\label{equation.exphotogluon}
       - 2\,e\,{\mathop{\rm Sin}\nolimits} 2{\theta _w}{K_{\gamma gg}}{\Theta _3}((\mu ,q),(\nu ,p),(\rho
       ,p')){\delta ^{ab}},
\end{equation}
where ${K_{\gamma gg}}$ is coupling constant of theory and we can assign it three numerical values: -0.098,
0.197 and -0.396~\cite{behr-TGB1,behr-TGB2}. In Eq.(~\ref{equation.exphotogluon}), $\Theta _3$ is given by:

\[\begin{array}{l}
{\Theta _3}((\mu ,{k_1}),(\nu ,{k_2}),(\rho ,{k_3})) = \\
 - ({k_1}.\theta .{k_2})\left[ {{{({k_1} - {k_2})}^\rho }{g^{\mu \nu }} + {{({k_2} - {k_3})}^\mu }{g^{\nu
 \rho }} + {{({k_3} - {k_1})}^\nu }{g^{\rho \mu }}} \right]\\
 - {\theta ^{\mu \nu }}\left[ {k_1^\rho ({k_2}.\,{k_3}) - k_2^\rho ({k_1}.\,{k_3})} \right] - {\theta ^{\nu
 \rho }}\left[ {k_2^\mu ({k_3}.\,{k_1}) - k_3^\mu ({k_2}.\,{k_1})} \right]\\
 - {\theta ^{\rho \mu }}\left[ {k_3^\nu ({k_1}.\,{k_2}) - k_1^\nu ({k_3}.\,{k_2})} \right] + ({\theta ^\mu
 }.\,{k_2})\left[ {{g^{\nu \rho }}k_3^2 - k_3^\nu k_3^\rho } \right]\\
 + ({\theta ^\mu }.\,{k_3})\left[ {{g^{\nu \rho }}k_2^2 - k_2^\nu k_2^\rho } \right] + ({\theta ^\nu
 }.\,{k_3})\left[ {{g^{\mu \rho }}k_1^2 - k_1^\mu k_1^\rho } \right]\\
 + ({\theta ^\nu }.\,{k_1})\left[ {{g^{\mu \rho }}k_3^2 - k_3^\mu k_3^\rho } \right] + ({\theta ^\rho
 }.\,{k_1})\left[ {{g^{\mu \nu }}k_2^2 - k_2^\mu k_2^\nu } \right] + ({\theta ^\rho }.\,{k_2})\left[ {{g^{\mu
 \nu }}k_1^2 - k_1^\mu k_1^\nu } \right]
\end{array}\]

 which is called the three-gauge boson vertex function~\cite{melic-strong}.

\end{enumerate}

\item In unexpanded approach~\cite{haghighat}:

\begin{enumerate}

\item Photon-fermion vertex is represented as:

\begin{equation}\label{equation.photoelec}
     - ie\;\exp (\frac{i}{2}k.\theta .q){\gamma ^\mu },
\end{equation}

\item Photon-gluon vertex has the following representation:

\begin{eqnarray}
&& e\;{\delta _{ab}}\,\Re \underbrace {({g_{\mu \nu }}{{(q - p)}_\rho } + {g_{\nu \rho }}{{(p + p')}_\mu } -
{g_{\rho \mu }}{{(p' + q)}_\nu })}_{{I^{\mu \nu \rho }}}\nonumber
 \\
&&\Re =  - \frac{2}{3}\sin (\frac{1}{2}q.\theta .p)\;. \label{eqnarray.unphotogluon}
\end{eqnarray}

\end{enumerate}

\end{itemize}
 Considering the condition ${\theta ^{\mu 0}} = 0$ the following useful identities will be obtained:
\begin{equation}\label{equation.iden1}
    A.\theta .B \equiv {A_\mu }{\theta ^{\mu \nu }}{B_\nu } = \vec \theta .\,(\vec A \times \vec B),
\end{equation}

\begin{equation}\label{equation.iden2}
    A.\,\theta .\,\theta .\,B = {A_\mu }{\theta ^{\mu \nu }}\theta _\nu ^\beta {B_\beta } = {\left| {\vec \theta
    } \right|^2}(\vec A.\,\vec B) - (\vec A.\,\vec \theta )(\vec B.\vec \theta ).
\end{equation}


\section{Electron-proton scattering in noncommutative space-tame}\label{section.eN}

Deep inelastic electron-proton scattering is a prevalent method to probe the proton.
Electron-proton cross section in laboratory system is given by~\cite{GreinerHalzen1,GreinerHalzen2}:
\begin{equation}\label{equation.crosssection}
\begin{array}{l}
\frac{{d\sigma }}{{d{Q^2}\,d\nu }} = \frac{{{\alpha ^2}\pi }}{{4{E^2}\,{{\sin }^4}(\varphi
/2)}}\frac{1}{{EE'}}\\
\quad \quad \quad \quad  \times [2{W_1}\,{\sin ^2}(\varphi /2) + {W_2}\,{\cos ^2}(\varphi /2)],
\end{array}
\end{equation}
where $\varphi$ , E and $E'$ are scattering angle and the energy of incident and scattered electrons,
respectively. The $W_1$ and $W_2$ functions characterize the structure of proton. In electron-parton elastic
scattering, partons (quarks and gluons) are assumed point like particles. To determine structure functions, usual
method is to consider electron which is scattered by quarks. For this purpose one can calculate the partonic
cross section. The result would be multiplied by parton distributions. Finally we need to integral over the
momentum fraction of each parton. In this paper, we use this method in our calculations. In the following we will
calculate electron-parton scattering in two approaches of NCFT where both the photon-quark and photon-gluon
interactions are considered.
\subsection{Parton model in expanded approach of NCSM}\label{subsection.sw}
 In NCSM, we follow the same method as in usual space-time with this exception that electron-quark scattering is
 corrected in NCST and additionally we consider as well the scattering of electron-gluon in our calculations. So
 we should take into account two individual contributions which we referred them before.\\
\textbf{Corrected vertex contribution:}  At first, we calculate electron-quark scattering with respect to given
vertex in Eq.(~\ref{eqnarray.photonfermion}).  In laboratory system, corrected vertex could be written as:

\begin{equation}\label{equation.eqlab}
    i\,{e_i}{g_e}{\gamma _\mu } + \frac{1}{2}{e_i}{g_e}\underbrace {\left[ { - (p'.\,{\theta _\mu })(\not p - m)}
    \right]}_{{C_\mu }}
\end{equation}
where $e_i$ is the charge of $i$th quark. After doing some algebraic task (see the Appendix), the average of
squared invariant amplitude for Electron-quark scattering in expanded approach is obtained as it follows:
\begin{equation}\label{equation.eq.invamp}
    \begin{array}{l}
\left\langle {{{\left| \cal{M}  \right|}^2}} \right\rangle  = \frac{1}{4}{(\frac{{{e_i}g_e^2}}{{{q^2}}})^2}
Tr[\,{\gamma ^\mu }\,(\not k + {m_e})\,{\gamma ^\nu }(\not k' + {m_e})]\;\\
\qquad\qquad\qquad Tr[\,{\gamma _\mu }\,(\not p + {m_q})\,{\gamma _\nu }(\not p' + {m_q})]\\
\qquad\qquad - \frac{i}{8}{(\frac{{{e_i}g_e^2}}{{{q^2}}})^2}Tr[\,{\gamma ^\mu }\,(\not k + {m_e})\,{\gamma ^\nu
}(\not k' + {m_e})]\;\\
\qquad\qquad\qquad Tr[\,{\gamma _\mu }\,(\not p + {m_q})\,{C_\nu }(\not p' + {m_q})]\\
\qquad \qquad - \frac{i}{8}{(\frac{{{e_i}g_e^2}}{{{q^2}}})^2}Tr[\,{\gamma ^\mu }\,(\not k + {m_e})\,{\gamma ^\nu
}(\not k' + {m_e})]\\
\qquad\qquad\qquad Tr[\,{C_\mu }\,(\not p + {m_q})\,{\gamma _\nu }(\not p' + {m_q})].
\end{array}
\end{equation}
In this equation the first term is corresponding to what is resulted from calculation in usual space-time and
relic terms arising from NCST. One can easily indicate by trace theorem that terms contain  NC parameter would be
vanished. Therefore, in this case, nucleon structure functions do not gain any correction from NCST and we will
have:
\begin{equation}\label{equation.structure function1}
    M{W_1}({Q^2},\nu ) = \frac{1}{2}\sum\limits_i {e_i^2\,{q_i}(x)}  \equiv {F_1}(x)
\end{equation}
\begin{equation}\label{eqution.structure function2}
    \nu {W_2}({Q^2},\nu ) = \sum\limits_i {e_i^2\,x\,{q_i}(x)}  \equiv {F_2}(x).
\end{equation}

\textbf{New vertex contribution:} Now we consider electron-gluon scattering in expanded NCST. Photon-gluon vertex
(see Eq.(~\ref{equation.exphotogluon})), considering the $N =  - 2\,\,{\mathop{\rm Sin}\nolimits} 2{\theta
_w}{K_{\gamma gg}}$ is written by:
\begin{figure}[t]
   \centerline{ \includegraphics[width=4cm]{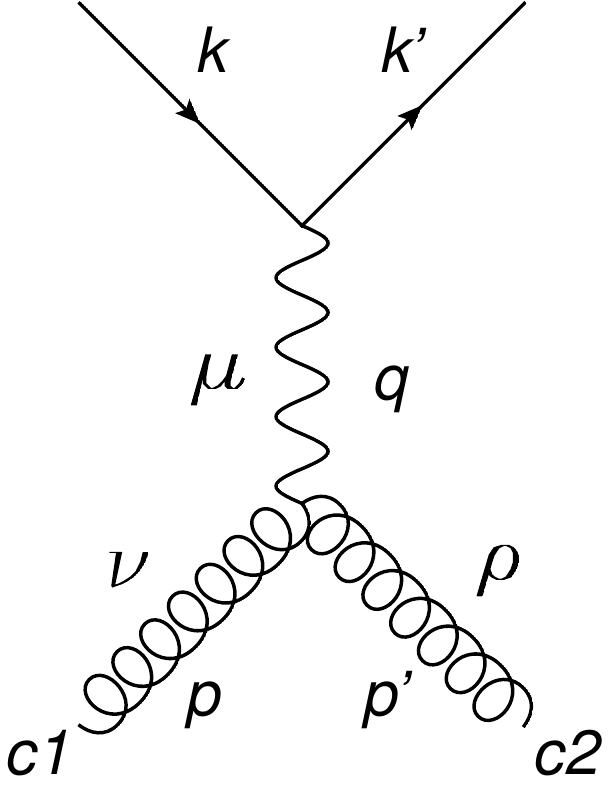}}
   \caption{A schematic graph for electron-gluon scattering.}
  \label{figure.eg}
\end{figure}
\begin{equation}\label{equation.eglab}
    {g_e}N\,\Theta {\delta ^{ab}}
\end{equation}
where for simplification, we omitted index 3 in $\Theta$. In the laboratory system and using identity, given by
Eq.(~\ref{equation.iden1}) the $\Theta$ quantity can be written as
\begin{equation}\label{equation.theta}
    \begin{array}{l}
\Theta ((\mu ,q),(\nu ,p),(\rho ,p')) = \\
\qquad - \theta ^{\mu \nu }\left[  - {q^\rho }({p^2}+ q.p)+ {p^\rho }({q^2} + q.p) \right]\\
  - {\theta ^{\nu \rho }}\left[ { - {p^\mu }{q^2} + {q^\mu }(q.p)} \right] + {\theta ^{\rho \mu }}\left[ {{p^\nu
  }(q.p) - {q^\nu }{p^2}} \right]\\
  - ({\theta ^\mu }.\,q)\left[ {{g^{\nu \rho }}{p^2} - {p^\nu }{p^\rho }} \right]
 + ({\theta ^\rho }.\,q)\left[ {{g^{\mu \nu }}{p^2} - {p^\mu }{p^\nu }} \right] \\
 + ({\theta ^\nu }.\,q)\left[ {{g^{\mu \rho }}({p^2} + 2q.p) - ({p^\mu }{p^\rho } + {p^\mu }{q^\rho } + {q^\mu
 }{p^\rho })} \right].
\end{array}
\end{equation}
Considering figure~\ref{figure.eg}, the invariant amplitude for electron-gluon scattering can be calculated. By
substituting gluon vertex (see Eq.(~\ref{equation.eglab})) in the expression for invariant amplitude, we will
have:
\begin{equation}\label{eqnarray.invamp.eg}
     {-i\cal{M}} = [\bar u(k')(i{g_e}{\gamma ^\lambda })u(k)][\frac{{ - i{g_{\lambda \mu
     }}}}{{{q^2}}}][{\varepsilon^* _{2\nu }}{c_{1}^ * }\,\,{g_e}N\,{\Theta ^{\mu \nu \rho }}\,{\delta
     ^{c_{1}c_{2}}}{\varepsilon _{3\rho }}c_{2}]
\end{equation}
where $\varepsilon$'s are gluon polarizations and $c_i$ is color factor of gluon. The symbol $\delta$ implies
that color changing is not happening for gluon. This is due to this fact the in photon-gluon interaction, photon
is a colorless identity. Nevertheless we should take into account contributions of each glouns since
the gluons can be appeared in eight color states.

Following the required calculations the corrected parts of structure function can be obtained (see the
Appendix):
\begin{equation}\label{equation.structure function eg1}
    M{W_1}({Q^2},\nu ) = \frac{{{M^2}b(x)}}{{{Q^2}}}x\,g(x) \equiv {F_1}(x),
\end{equation}
\begin{equation}\label{equation.structure function eg2}
    \nu {W_2}({Q^2},\nu ) = a(x)\,x\,g(x) \equiv {F_2}(x),
\end{equation}
where
\begin{equation}\label{equation.coff.a}
   \begin{array}{l}
a(x) = \frac{{{N^2}{\theta ^2}}}{{2}}\left( - 12xM{E^3} - 6{x^2}{M^2}{E^2} + 12{E^2}{Q^2}\right.\\
\qquad - 5{x^2}{M^2}{Q^2}  + 40xM{E^2}E' + 16{x^2}{M^2}EE'\\
\qquad - 22EE'{Q^2} - 40xME{{E'}^2} - 6{x^2}{M^2}{{E'}^2}\\
\qquad\qquad\left.  + 12{{E'}^2}{Q^2} + 12xM{{E'}^3}+ 6{Q^4}\right)
\end{array}
\end{equation}
and
\begin{equation}\label{equation.coff.b}
    \begin{array}{l}
b(x) = \frac{{{N^2}{\theta ^2}}}{{2}}( - 8{E^4} - 8{{E'}^4} + 4xM{E^3} + 2{x^2}{M^2}{E^2}\\
\quad \quad + 24xME{{E'}^2} + 4{x^2}{M^2}{Q^2} - 24xM{E^2}E'\\
\quad \quad  - 6{x^2}{M^2}EE'- 4EE'{Q^2}  + 2{x^2}{M^2}{{E'}^2}\\
\quad \quad  - 4xM{{E'}^3} + \frac{{11{Q^4}}}{4}).
\end{array}
\end{equation}
Here $Q^2=-q^2$ and $q$ is the transferred momentum by photon, M is mass of proton and other parameters are
defined by:
\begin{equation}\label{equation.nu}
  \nu  = \frac{{{Q^2}}}{{2Mx}}, \end{equation}
\begin{equation}\label{equation.E}
    E = \frac{\nu }{y} = \frac{{{Q^2}}}{{2Mxy}},   \end{equation}
\begin{equation}\label{equation.E'}
   E' = E - \frac{{{Q^2}}}{{2Mx}}.  \end{equation}
The $\theta^2$ is squared of $\vec \theta $ and the energy scale($\Lambda_{NC}$) for NCST is given by:
\begin{equation}\label{equation.scaling}
\left| \vec \theta \right| = \frac{1}{{{\Lambda_{NC} ^2}}} \end{equation}
The final result for nucleon structure function would be obtained by adding the gluon effect to the rest of
contributions. Therefore we will get the following results:
\begin{equation}\label{equation.struct.total1}
    M{W_1}({Q^2},\nu ) = \frac{1}{2}\sum\limits_i {e_i^2\,{q_i}(x)}  + \frac{{{M^2}b(x)}}{{{Q^2}}}x\,g(x) \equiv
    {F_1}(x)
\end{equation}
\begin{equation}\label{equation.struct.total2}
    \nu {W_2}({Q^2},\nu ) = \sum\limits_i {e_i^2\,x\,{q_i}(x)}  + a(x)\,x\,g(x) \equiv {F_2}(x)
\end{equation}
where $q_i$ and $g_i$ are distribution functions of quarks and gluons, respectively.\\
In the following section, we will use Eq.(~\ref{equation.struct.total2}) to indicate the effect of gluon
distribution to modify the proton structure function, resulted from NC modification.
\subsection{Parton model in unexpanded approach of NCST}\label{subsection.cpst}
In unexpanded approach, calculations are like the ones in expanded approach except that we should use the vertex,
given by Eqs.(~\ref{equation.photoelec},~\ref{eqnarray.unphotogluon}).\\
\textbf{Corrected vertex contribution:} By replacing photon-electron corrected vertex into leptonic tensor,
$L^{\mu \nu }$, this tensor will be appeared as:
\begin{equation}\label{equation}
    \begin{array}{l}
{L^{\mu \nu }} = \frac{1}{2}\sum\limits_{\,Spins} {[\bar u(k'){\gamma ^\mu }{e^{\frac{i}{2}k.\theta
.q}}\;u(k)]{{[\bar u(k'){\gamma ^\nu }{e^{\frac{i}{2}k.\theta .q}}u(k)]}^ * }} \\
\quad \;\,\, = \frac{1}{2}\sum\limits_{\,Spins} {\bar u(k'){\gamma ^\mu }{e^{\frac{i}{2}k.\theta .q}}\;u(k)\,\bar
u(k){\gamma ^\nu }{e^{ - \frac{i}{2}k.\theta .q}}u(k')} \\
\quad \;\,\, = \frac{1}{2}\sum\limits_{\,Spins} {\bar u(k'){\gamma ^\mu }\;u(k)\bar u(k){\gamma ^\nu }u(k')}.
\end{array}
\end{equation}
It is obvious that no correction is arising from NCST in leptonic tensor. Consequently one can show as well that
there is not any correction in partonic sector.
\\
\textbf{New vertex contribution:} Starting from Eq.(~\ref{eqnarray.unphotogluon}), following the calculation
that listed in appendix for electron-gluon scattering in the expanded NC and using the definition $H =
\frac{4}{{9}} - \frac{2}{{9}}\frac{{{M^2}}}{{{Q^2}}}$, one obtains
\begin{equation}\label{equation.struct1.eg.unex}
    M{W_1}({Q^2},\nu ) = H\,g(x){\sin ^2}(\frac{1}{2}q.\theta .p) \equiv {F_1}(x),
\end{equation}
\begin{equation}\label{equation.struct2.eg.unex}
    \nu {W_2}({Q^2},\nu ) = \frac{5}{{9}}x\,g(x){\sin ^2}(\frac{1}{2}q.\theta .p) \equiv {F_2}(x).
\end{equation}
Since our calculations are in laboratory system and according to Eq.(~\ref{equation.iden1}), in this case, we
also do not have any gluon contribution, therefore, unexpanded approach of NC does not have any effect on nucleon
structure functions in laboratory system and consequently the structure functions would be as in usual space-time
which are given by Eqs.(~\ref{equation.structure function1},~\ref{eqution.structure function2}).

\section{Results and discussions}\label{section.result}
In section~\ref{section.eN} correction of NCST has been calculated up to leading order in terms of NC parameter,
$\theta$. NC correction on structure functions comes out from electron-gluon scattering in the expanded approach.
By writing Eq.(~\ref{equation.struct.total2}) in terms of constituent quarks and gluons distributions we will
have the following result for proton structure function:
\begin{equation}\label{equation.constituent}
    \begin{array}{l}
{F_2}(x) = {\left( {\frac{2}{3}} \right)^2}\left[ {x{u_v}(x) + 2x\,\bar u(x)} \right] + {\left( { - \frac{1}{3}}
\right)^2}\left[ x{d_v}(x)\right.\\
\left. + 2x\,\bar d(x) \right]+ {\left( { - \frac{1}{3}} \right)^2}\left[ {2x\,s(x)} \right] + a(x)\,x\,g(x).
\end{array}
\end{equation}
where $u(x), d(x), s(x)$ and $g(x)$ denote the quarks and gluon distribution functions. The final term comes from
our calculations in NCST. Factor $a(x)$ (see Eq.(~\ref{equation.coff.a})) contains parameters of NCST like
$\theta$ and $K_{\gamma gg}$, and usual parameters like the energies of the incident and scattered electron ($E$
and $E'$), transferred momentum($q$ as $Q^2=-q^2$), proton mass ($M$) and the momentum fraction carried by each
parton ($x$).

\begin{figure}[hb]
\begin{center}
\includegraphics[width=8cm]{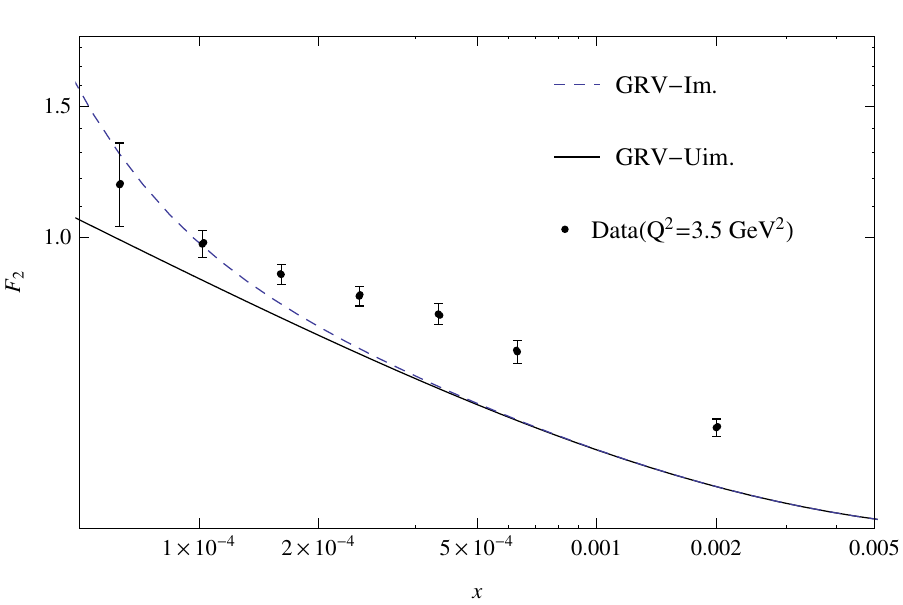}
\includegraphics[width=8cm]{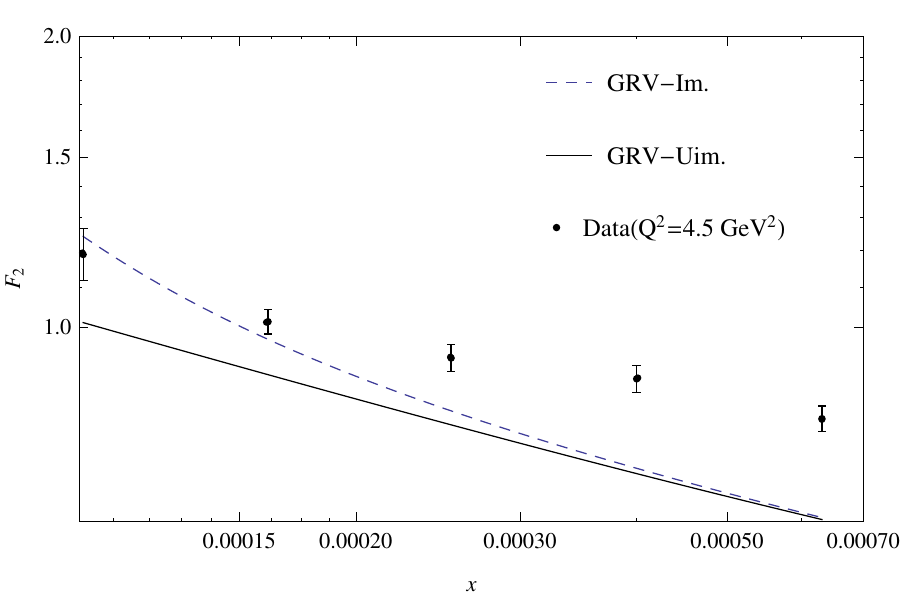}
\includegraphics[width=8cm]{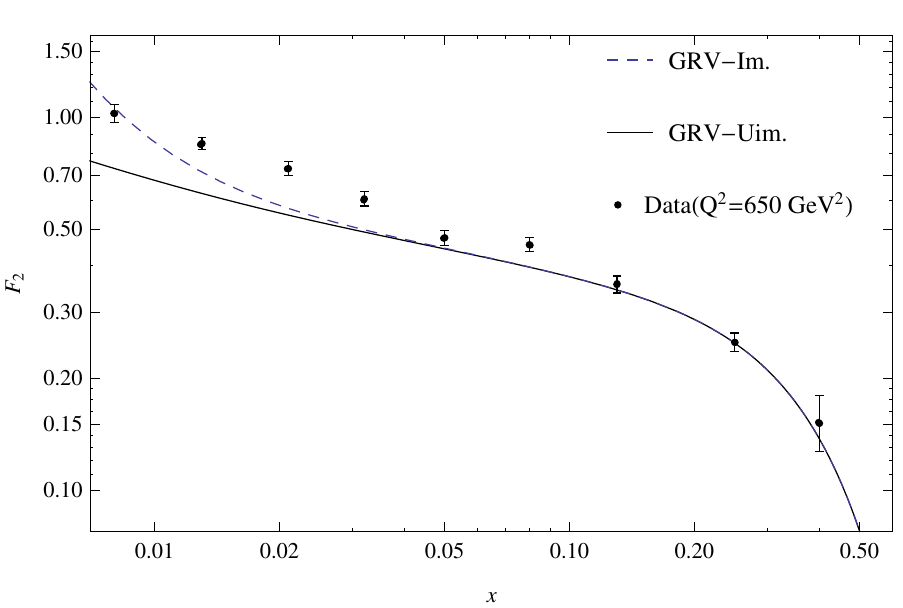}
\end{center}
\caption{\label{figH}{Our results for the modified  nucleon structure function(NSF) , $F_2(x)$, at
$Q^2=3.5\;GeV^2$, $4.5\;GeV^2$ and $ 650\;GeV^2$ which are compared with the available experimental data
~\cite{data} and the normal GRV parametrization model ~\cite{GRV}. Here the "GRV-Im."  is indicating our results
for the modified NSF, using GRV model. The "GRV-Uim." symbol  is denoting the normal NSF in GRV model.}}
\end{figure}

\begin{figure}[hb]
\begin{center}
\includegraphics[width=8cm]{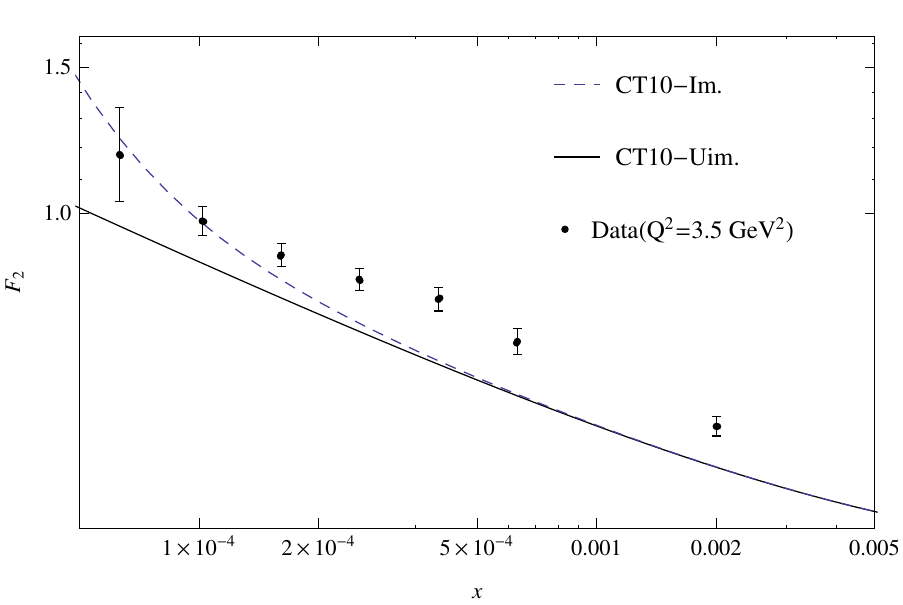}
\includegraphics[width=8cm]{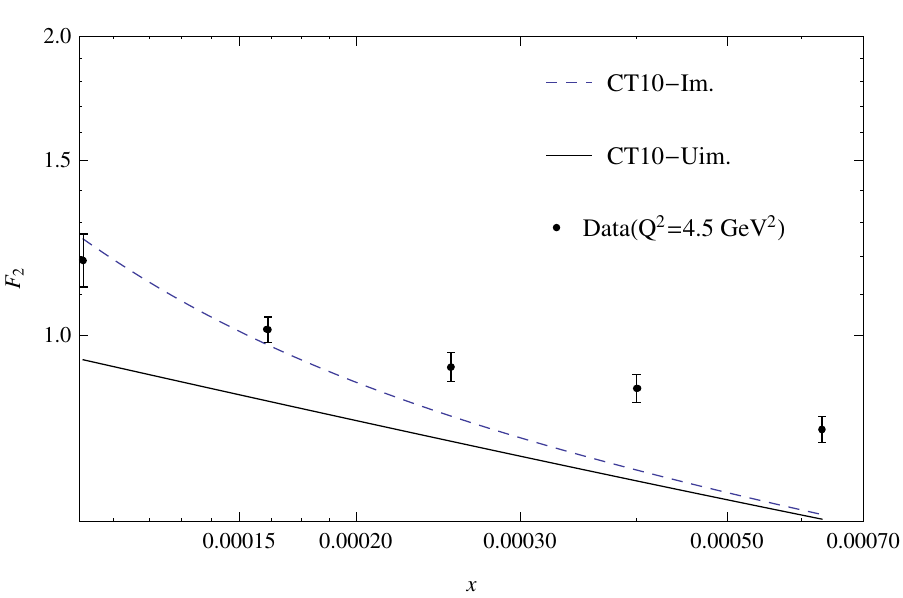}
\includegraphics[width=8cm]{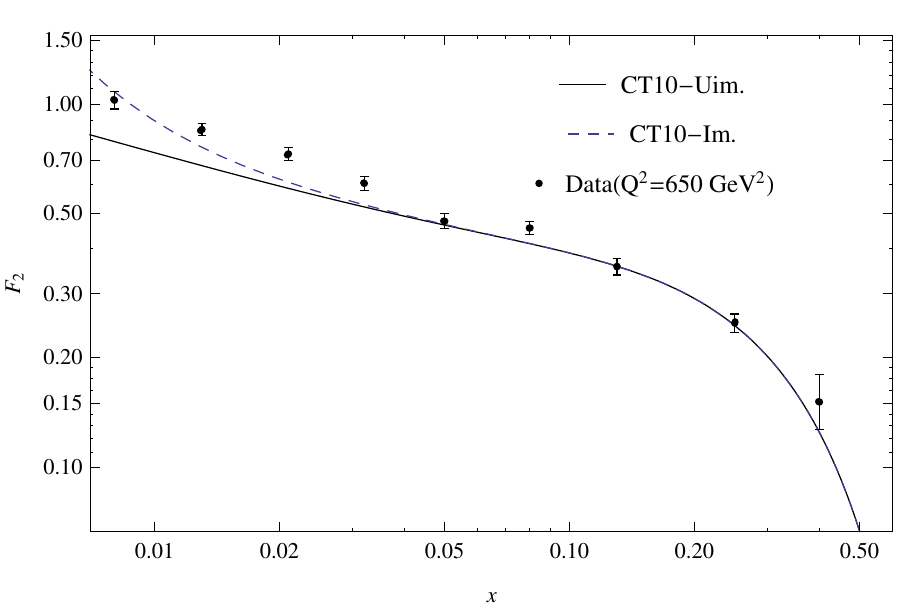}
\end{center}
\caption{{\label{figH1}Our results for the modified  nucleon structure function(NSF) , $F_2(x)$, at
$Q^2=3.5\;GeV^2$, $4.5\;GeV^2$ and $ 650\;GeV^2$ which are compared with the available experimental data
~\cite{data} and the normal CT10 parametrization model ~\cite{CT10}. Here the "CT10-Im."  is indicating our
results for the modified NSF, using CT10 model. The "CT10-Uim." symbol  is denoting the normal NSF in CT10
model.}}
\end{figure}

\begin{figure}[!hbp]
\begin{center}
\includegraphics[width=8cm]{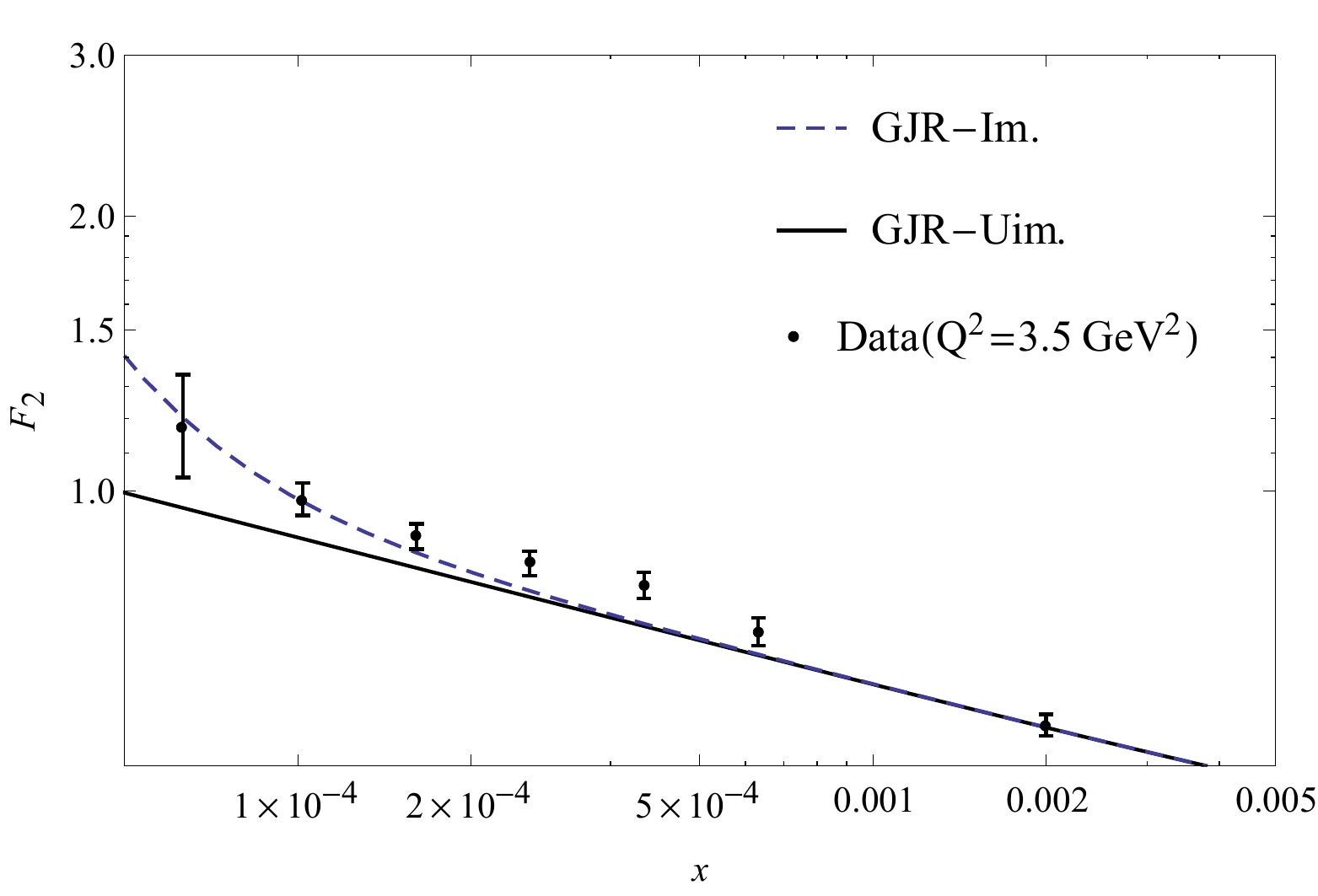}
\includegraphics[width=8cm]{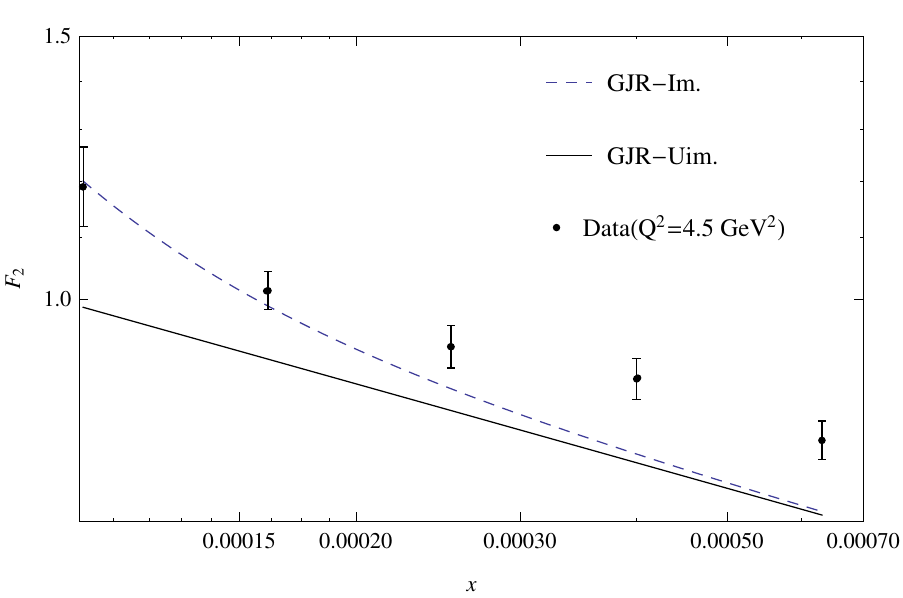}
\includegraphics[width=8cm]{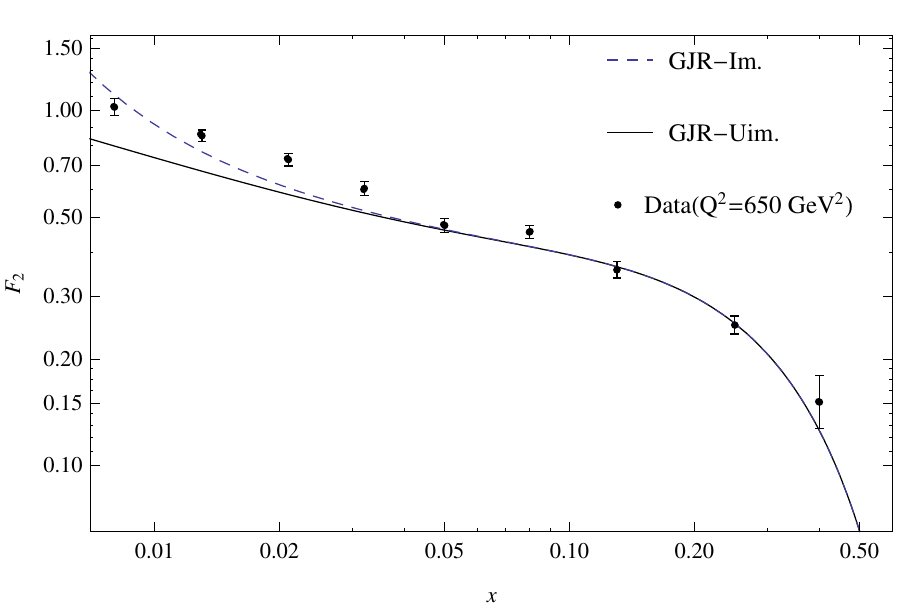}
\end{center}
\caption{{\label{figH2}Our results for the modified  nucleon structure function(NSF) , $F_2(x)$, at
$Q^2=3.5\;GeV^2$, $4.5\;GeV^2$ and $ 650\;GeV^2$ which are compared with the available experimental data~\cite{data} and the normal GJR(JR09FFNNLO) parametrization model~\cite{GJR}. Here the "GJR(JR09FFNNLO)-Im."  is
indicating our results for the modified NSF, using GJR model. The "GJR(JR09FFNNLO)-Uim." symbol  is denoting the
normal NSF in CT10 model.}}
\end{figure}

\begin{figure}[!hbp]
\begin{center}
\includegraphics[width=8cm]{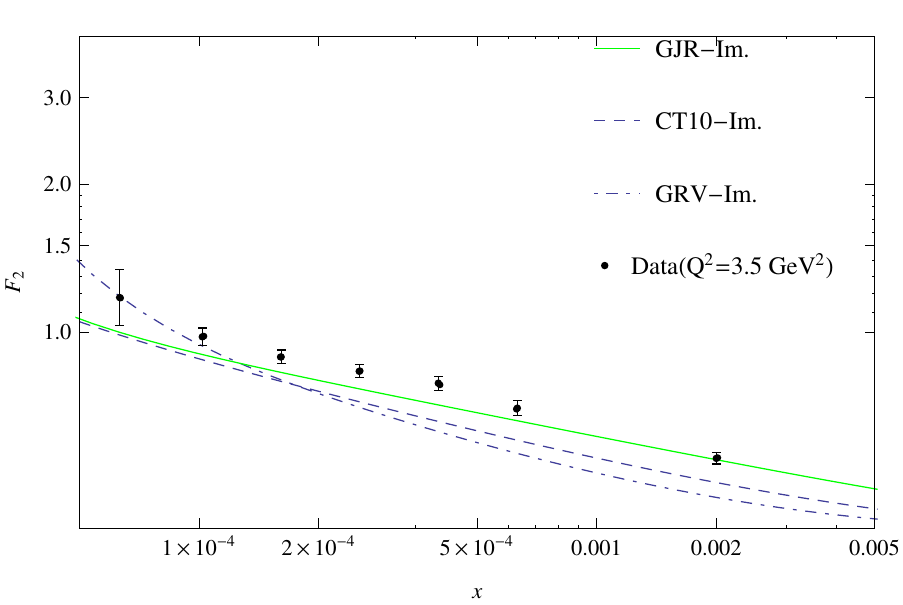}
\includegraphics[width=8cm]{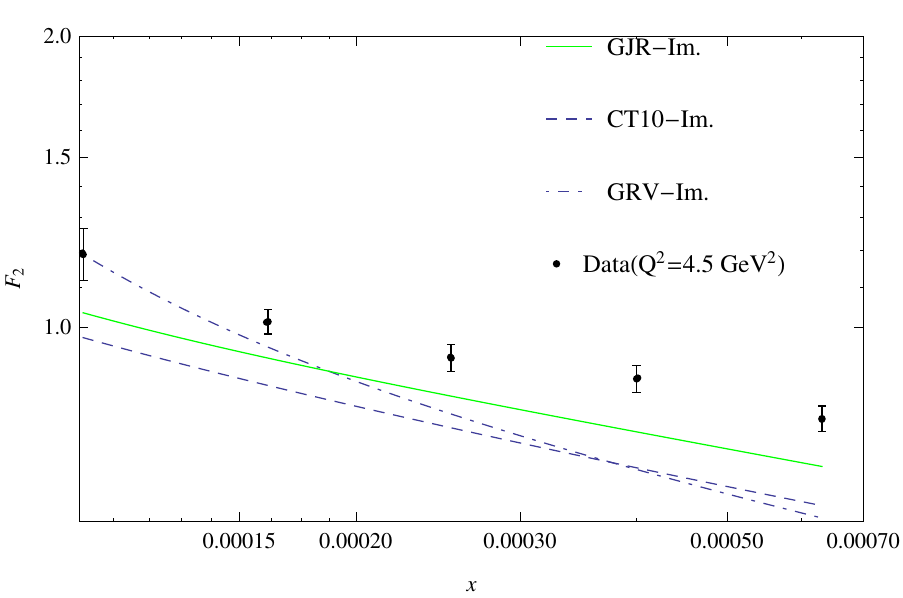}
\includegraphics[width=8cm]{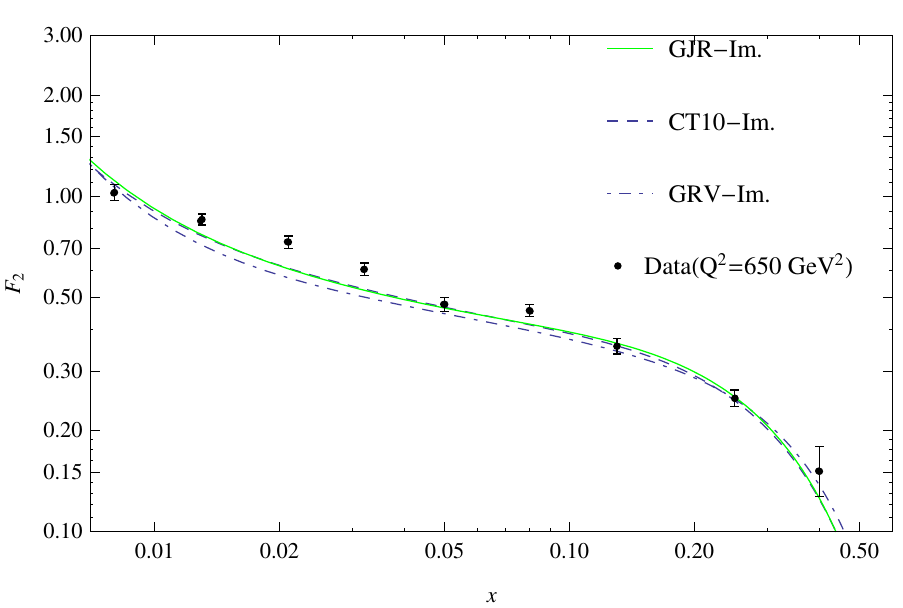}
\end{center}
\caption{{\label{figH3} Our results for the modified  nucleon structure function(NSF) , $F_2(x)$, at
$Q^2=3.5\;GeV^2$, $4.5\;GeV^2$ and $ 650\;GeV^2$ which are compared with the available experimental data~\cite{data}, using GRV, CT10 and  GJR(JR09NNLO) models. }}
\end{figure}

We have depicted the modified nucleon  structure function ($F_2(x)$) in figure~\ref{figH}
by substituting Eqs (~\ref{equation.nu}), (~\ref{equation.E}) and (~\ref{equation.E'}) in Eq
(~\ref{equation.coff.a}).
The value of $Q^2$ and $y$ have been chosen to correspond  the available range of experimental date.
The results have been compared with available experimental data~\cite{data} and
the prediction of GRV parametrization model~\cite{GRV}.
To indicate the theoretical uncertainty in the  standard model prediction for the structure function $F_2(x)$, we
also use the GJR and CT10  parametrization models~\cite{GJR,CT10} and make the results for the  modified nucleon
structure function . In  figures~\ref{figH1},~\ref{figH2} we depict the results for these two models in the
modified and normal cases. A comparison with the available experimental  data has also been done. In figure~\ref{figH3} the results for the nucleon structure function, arising from the modified models are compared with
each other and also with  the available experimental data. As we are expecting, the theoretical uncertainty,
using different parametrization models is very low and we get a firm conclusion for the validity of the modified
models, considering the NC effect.

To plot the $F_2(x)$ we need two NC parameters, $\Lambda_{NC}$ and $K_{\gamma gg}$. 
good fits  for $F_2$ with experimental date is obtained for approved amounts of energy scale $\Lambda_{NC}$.
As it has been mentioned above, looking for the literature and concerned articles, there is not specific value
for the NC scale. Collider scattering experiments could be a proper evident to search NCST effects because they
are very sensitive  to NC signals. Usual bound from these experiments is $\Lambda_{NC}\sim 1TeV $.  Present work
is also  implemented for collider scattering experiment to find modified structure functions of nucleons in NCST.
So we have also employed this range that is prevalent in such processes. \\

Following the procedure which was described in the expanded approach, one can find three numerical values for
the $K_{\gamma gg}$ parameter which are  -0.098, 0.197 and -0.396 respectively~\cite{behr-TGB1,behr-TGB2}.
$\Lambda_{NC}$. Therefore we investigate these effects.  The results  for all three values of $K_{\gamma gg}$ are
similar to each other. Therefore we just present the results coming from the numerical values of parameters and
scales which are tabulated in table~\ref{table.I}.
As can be seen from table~\ref{table.I}, the numerical value for NC scale is growing by increasing the $Q^2$ as
squared transfer momentum. For example for a fixed  $K_{\gamma gg}$, the  $\Lambda_{NC}= 800\; GeV$ for
$Q^2=3.5\; GeV^2$  while  for  $Q^2=650\; GeV^2$ the $\Lambda_{NC}$ scale is equal to $ 2200\; GeV$.  It is in
correspond  to our expectation from the NC effect.
The $\Lambda_{NC}$ also depends on the measure of  $K_{\gamma gg}$ parameter. According to table~\ref{table.II}
for a fixed $Q^2$, the value of $\Lambda_{NC}$ increases when the magnitude of $K_{\gamma gg}$ is coming up. For
instance, at a fixed $Q^2$ for $K_{\gamma gg}=-0.098$ we will get $\Lambda_{NC}=2200\; GeV$  and it is $4400\;
GeV$ for  $K_{\gamma gg}=-0.396$.   The numerical values for NC scale, using the GJR and CT10 parametrization
models are at the same order of the ones in tables~\ref{table.I} and~\ref{table.II} with similar behaviour.%
when the squared transfer momentum is increasing.
\begin{table}[tbp]
\centering
\begin{tabular}{|c|c|c|}
\hline
 $Q^2 (GeV)^2 $ &$K_{\gamma gg}$ & $\Lambda_{NC} (GeV) $  \\
\hline
 $3.5$ & $-0.098$ &  $830$ \\
\hline
 $ 4.5 $ &  $-0.098$  &  $750$  \\
\hline
 $ 650$  &  $-0.098$  &  $2200$  \\
\hline
\end{tabular}
\caption{\label{table.I} $\Lambda_{NC}$ variations with squared transfer momentum for a fixed $ K_{\gamma gg} =
-0.098$   
}
\end{table}
\begin{table}[tbp]
\centering
\begin{tabular}{|c|c|c|}
\hline
$Q^2$ $ (GeV)^2 $ & \qquad $ K_{\gamma gg} $ \qquad & $\Lambda_{NC} (GeV) $ \\
\hline
  $ 650$  &  $-0.098$  &  $2200 $\\
\hline
  $650 $ &  $+0.197$  &  $3200$ \\
\hline
  $650$  &  $-0.396$  &  $4400$ \\
\hline
\end{tabular}
\caption{\label{table.II} $\Lambda_{NC}$ variations with parameter $ K_{\gamma gg} $ for a fixed $Q^2=650 (GeV)^2
$
}
\end{table}
Figures~\ref{figH},~\ref{figH1} and~\ref{figH2} indicate good compatibility with experimental data especially
for small values of $x$-Bjorken variable where we are expecting the effect of new physics are more relevant. To
confirm the validity of our obtained results we can act conversely and concentrate to extract the energy scale,
$\Lambda_{NC}$. We then need to consider Eq.(~\ref{equation.constituent}) while the energy scale, $\Lambda_{NC}$
is unknown. If, at the fixed $K_{\gamma gg}$, we use  experimental data at low $Q^2$, a small value for $\Lambda_{NC}$ is obtained and for data at the highest   $Q^2$  value a large value for $\Lambda_{NC}$ would be appeared. What we get are as following:

\begin{itemize}
\item For $K_{\gamma gg}=-0.098$:\qquad $\Lambda_{NC}\geq 430\; GeV$
\item  For $K_{\gamma gg}=0.197$:\qquad\ \  $\Lambda_{NC} \geq 610\; GeV$
\item  For $K_{\gamma gg}=-0.396$:\qquad $\Lambda_{NC} \geq 860\; GeV$
\end{itemize}
However  in obtaining the above numerical values for $\Lambda_{NC}$ scale we use the GRV model but similar
results will be appeared when we employ the GJR and CT10 models.
To do a confirmation on the validity of our calculations, we once again back to the  Drell-Yan process which is
an important role for investigating the nucleon structure function and in testing the parton model. Analysing
this processes at the NCST will yield us $\Lambda_{NC}\geq 0.4\; TeV$~\cite{phenomenological12} which is
compatible with what we get in this paper.
\section{Conclusion}\label{section.conclusion}
We have considered the effect of NCST on proton structure functions. There are two approaches to construct the
usable NC theory. In both approaches, all present vertices are modified by the NC parameter $\theta_{\mu\nu}$. In
this case, in addition to the usual interactions, some new interactions would also be appeared. We have applied
two new corrections and  two new interactions, one for each approach, to calculate the structure functions of
proton. Three of the four corrections do not have any effect, but new interaction from expanded approach
contributes to the nucleon structure function.
As can be seen, the obtained results for the improved proton structure function, $F_2(x)$ , are in better
compatibility with available  experimental data rather than the results coming from  the normal GRV, CT10 and GJR
parametrization models, specially at small values of $x$-Bjorken variable which is related to the high energy
region.
Also the magnitude order of NC energy scale that we got, using NCSM approach, is correspond to the expected range
of the other predictions.

In this paper we considered a spacial case when $\theta_{\mu 0}=0$  but one can calculate the case for
$\theta_{\mu 0}\neq0$ which we hope to report about it in future. The current results can be extended to the
higher order approximation which we hope to do it as our new research task. The NCSM which we used it here,
contained the lorentz violation. Some similar calculations in which lorentz invariant is conserved can be done
which we hope to report on this issue latter on.
\appendix
\section{Appendix}
Here, we perform the required calculations for electron-quark and electron-gluon scattering in the expanded NC in
details. Similar calculations can be done for unexpanded NC.

\textbf{Electron-quark scattering:} Employing Feynman rule for figure~\ref{figure.eg} we will able to obtain the
required results up to leading order, considering the NC parameter. Since propagators do not affected by NC
corrections therefore just vertexes should be written in NC space-time. According to photon-fermion vertex in
laboratory system (see Eq.(~\ref{equation.eqlab})), invariant amplitude read as it follow:
\begin{equation}\label{equation.invariant amp. eq}
   \begin{array}{l}
  {-i\cal{M}} = [\bar u(k')(i{g_e}{\gamma ^\mu })u(k)][\frac{{ - i{g_{\mu \nu }}}}{{{q^2}}}][\bar
  u(p')(i{e_i}{g_e}{\gamma ^\nu })u(p)]\\
\quad \quad \;\; + [\bar u(k')(i{g_e}{\gamma ^\mu })u(k)][\frac{{ - i{g_{\mu \nu }}}}{{{q^2}}}][\bar
u(p')(\frac{1}{2}{e_i}{g_e}{C^\nu })u(p)].
\end{array}
\end{equation}
Doing some simplification we will have:
\begin{equation}\label{equation.invar.amp.eq.simp}
    \begin{array}{l}
 {-\cal{M}}  = \frac{{{e_i}g_e^2}}{{{q^2}}}[\bar u(k'){\gamma ^\mu }u(k)][\bar u(p'){\gamma _\mu }u(p)]\\
\quad \quad  - i\frac{{{e_i}g_e^2}}{{2{q^2}}}[\bar u(k'){\gamma ^\mu }u(k)][\bar u(p'){C_\mu }u(p)].
\end{array}
\end{equation}
Then for the squared invariant amplitude, we will get:
\begin{equation}\label{equation.inv.amp.square}
    \begin{array}{l}
{{\left| \cal{M}  \right|^2}} = {(\frac{{{e_i}g_e^2}}{{{q^2}}})^2}[\bar u(k'){\gamma ^\mu }u(k)][\bar
u(p'){\gamma _\mu }u(p)]{[\bar u(k'){\gamma ^\nu }u(k)]^\dag }\\
{[\bar u(p'){\gamma _\nu }u(p)]^\dag }+ \frac{i}{2}{(\frac{{{e_i}g_e^2}}{{{q^2}}})^2}[\bar u(k'){\gamma ^\mu
}u(k)][\bar u(p'){\gamma _\mu }u(p)]\\
{[\bar u(k'){\gamma ^\nu }u(k)]^\dag }{[\bar u(p'){C_\nu }u(p)]^\dag } -
\frac{i}{2}{(\frac{{{e_i}g_e^2}}{{{q^2}}})^2}{[\bar u(k'){\gamma ^\mu }u(k)]}\\
{[\bar u(p'){C_\mu }u(p)]}{[\bar u(k'){\gamma ^\nu }u(k)]^\dag }{[\bar u(p'){\gamma _\nu }u(p)]^\dag }.
\end{array}
\end{equation}
Here, we remember that for two $4\times4$ ${\Gamma _1}$ and ${\Gamma _2}$ matrices, Casimir's trick will lead us
to:
\begin{equation}\label{equation. casimir trick}
\begin{array}{l}
  \sum\limits_{all\;spins} {[\bar u(a){\Gamma _1}u(b)]{{[\bar u(a){\Gamma _2}u(b)]}^\dag }} = \\
  \qquad\qquad\qquad Tr[\,{\Gamma _1}\,({\not p_b} + {m_b})\,{\bar \Gamma _2}({\not p_a} + {m_a})]\;.
\end{array}
\end{equation}
According to ${\bar \Gamma _2} = {\gamma ^0}\Gamma _2^\dag \,{\gamma ^0}$ definition, we have:
${\gamma ^0}{C^{\nu \dag }}\,{\gamma ^0} =  - {C^\nu }$, ${\gamma ^0}{\gamma ^{\nu \dag }}\,{\gamma ^0} = {\gamma
^\nu }$. Now by taking average over initial spin states and sum over final spin states and using the Casimir's
trick we arrive at Eq.(~\ref{equation.eq.invamp}).

\textbf{Electron-gluon scattering:} To do  the required calculations, we consider figure~\ref{figure.eg} and
proceed to do the square of invariant amplitude in Eq.(~\ref{eqnarray.invamp.eg}). Then doing the average over
initial spins states and sum over final spin states and then gluon polarization states so as:
\begin{equation}\label{equation.pol.gluon}
  \sum {{\varepsilon _\mu }\varepsilon _\nu ^ * }  \to  - {g_{\mu \nu }},
\end{equation}
and color algebra
\begin{equation}\label{color.algebra}
    \sum\limits_{{c_1}{c_2}} {{\delta ^{{c_1}{c_2}}}{\delta _{{c_1}{c_2}}}}  = \sum\limits_{{c_1} = 1}^8 {{\delta
    ^{{c_1}{c_1}}}}  = 8,
\end{equation}

we will get the following result:

\begin{equation}\label{equation.eg.invamp.long}
   \begin{array}{l}
\left\langle {{\left| \cal{M} \right|}^2} \right\rangle  =
\frac{16{N^2}g_e^4}{q^4}\left[\theta^2\left(k.k'\,(p^2+q^2)(-p^2 q^2+(q.p)^2)\right.\right.\\
\left.\left. -q^2q.p\,(k.pk'.q+k.qk'.p)+q^4 k.p\,k'.p+k.q\,k'.q\,(q.p)^2\right.\right.\\
\left.\left. -k.k'\, q.p\, (p^2 q^2-(q.p)^2)\right)-p^4 k.\theta .q\, k'.\theta .q+k.\theta.\theta.q \right.\\
\left.\times(q^2p^2\,k'.(q+p))+k.\theta.\theta. k'(p^2+(p+q)^2)(p^2q^2\right.\\
\left.-(p.q)^2)-q.\theta\theta.q\left(p^2\,k.k'(\frac{p^2}{2}+2q^2+3q.p)+k.pk'.p\right.\right.\\
\left.\left. (2p^2+q^2+2p.q)+(k.qk'.p+k.p\,k'.q)(p^2+2p.q)\right)\right.\\
\left.-p^2k.qk'.q  +k'.\theta.\theta. q \left((q^2\,p^2-2(q.p)^2)k.(p+q)\right)\right.\\
\left.\left.+(p^2\,k.q+q^2\,k.p)\,q.p\right)\right]
  \end{array}
\end{equation}

In Eq.(~\ref{equation.eg.invamp.long}) we neglected from electron mass. We note that since there is not more than
two gluon legs, thus incident gluon is like the outgoing gluon. Mathematically, the delta Kronecker function
confirms this reality. On the other hand since gluons are appearing in eight color states, we should consider the
color factor in our calulation which can be done, using Eq.({~\ref{color.algebra}}). Now to simplify the above
equation, by takeing $\alpha $, $\beta $, $\gamma $ as angles between $\vec k$, $\vec k'$, $\vec k \times \vec
k'$ and $\vec \theta $ direction, respectively, in the laboratory system and using Eqs.(~\ref{equation.iden1})
and (~\ref{equation.iden2}) we will get:
\begin{equation}\label{equation.alpha}
    k.\,\theta .\,\theta .\,k = {E^2}|\vec \theta {|^2}{\sin ^2}\alpha ,
\end{equation}
\begin{equation}\label{equation.beta}
    k'.\,\theta .\,\theta .\,k' = {E'^2}|\vec \theta {|^2}{\sin ^2}\beta ,
\end{equation}
\begin{equation}\label{equation.alpha.beta.phi}
    k.\,\theta .\,\theta .\,k' = E\,E'|\vec \theta {|^2}(\cos \varphi  - \cos \beta \cos \alpha ),
\end{equation}
\begin{equation}\label{equation.gama}
    k.\,\theta .\,k' = E\,E'|\vec \theta |\sin \varphi \cos \gamma \;.
\end{equation}
Then by taking the average over $\alpha $, $\beta $ and $\gamma $, Eq.(~\ref{equation.eg.invamp.long}) will lead
us to:
\begin{equation}\label{equation.c}
    \left\langle {{{\left| \cal{M} \right|}^2}} \right\rangle  = \frac{{8{N^2}g_e^4}}{{{q^4}}}{\theta
    ^2}EE'm_{eff}^2\left[ {a'{{\cos }^2}(\frac{\varphi }{2}) + b'{{\sin }^2}(\frac{\varphi }{2})} \right],
\end{equation}
where
\begin{equation}\label{equation.a'}
    \begin{array}{l}
a' =  - 12{m_{eff}}{E^3} - 6m_{eff}^2{E^2} + 12{E^2}{Q^2} - 5m_{eff}^2{Q^2} \\
\quad +6{Q^4} + 40{m_{eff}}{E^2}E' + 16m_{eff}^2EE' - 40{m_{eff}}E{{E'}^2}\\
\quad - 22EE'{Q^2}- 6m_{eff}^2{{E'}^2} + 12{{E'}^2}{Q^2} + 12{m_{eff}}{{E'}^3},
\end{array}
\end{equation}
\begin{equation}\label{equation.b'}
    \begin{array}{l}
b' =  - 8{E^4} - 8{{E'}^4} + 4{m_{eff}}{E^3} + 2m_{eff}^2{E^2}\\
\quad  + 4m_{eff}^2{Q^2} - 24{m_{eff}}{E^2}E' - 6m_{eff}^2EE'\\
\quad  - 4EE'{Q^2} + 24{m_{eff}}E{{E'}^2} + 2m_{eff}^2{{E'}^2}\\
\quad+ \frac{{11{Q^4}}}{4} - 4{m_{eff}}{{E'}^3}.
\end{array}
\end{equation}
Here ${m_{eff}}$ is zeroth component of four-momentum for gluon. By substitute Eq.(~\ref{equation.c}) into below
equation
\begin{equation}\label{equation. xxx}
    \begin{array}{l}
d\sigma  = \frac{1}{{(2E)(2m)}}\frac{{\left\langle {{{\left| \cal{M} \right|}^2}} \right\rangle }}{{4{\pi
^2}}}\frac{{{d^3}k'}}{{2E'}}\frac{{{d^3}p'}}{{2{{p'}_0}}}{\delta ^{(4)}}(p + k - p' - k')\\
\quad \;\; = \frac{1}{{4mE}}\frac{{\left\langle {{{\left| \cal{M} \right|}^2}} \right\rangle }}{{4{\pi
^2}}}\frac{1}{2}E'dE'd\Omega \frac{{{d^3}p'}}{{2{{p'}_0}}}{\delta ^{(4)}}(p + k - p' - k'),
\end{array}
\end{equation}
and using
\begin{equation}\label{equation.yyy}
    \int {\frac{{{d^3}p'}}{{2{{p'}_0}}}{\delta ^{(4)}}(p + q - p')}  = \frac{1}{{2m}}\delta (\nu  +
    \frac{{{q^2}}}{{2m}}),
\end{equation}
where in laboratory system we have
\begin{equation}\label{equation.driv}
    d(\cos \varphi )dE' = \frac{1}{{2EE'}}d{Q^2}d\nu,
\end{equation}
we obtain
\begin{equation}\label{equation.ori}
\begin{array}{l}
    \frac{{d\sigma }}{{d{Q^2}d\nu }} = \frac{{\pi {\alpha ^2}}}{{4{E^2}{{\sin }^4}(\frac{\varphi
    }{2})}}\frac{1}{{EE'}}\\
  \qquad\qquad  \left[ {a\,{{\cos }^2}(\frac{\varphi }{2}) + b{{\sin }^2}(\frac{\varphi }{2})} \right]\delta (\nu
  - \frac{{{Q^2}}}{{2{m_{eff}}}}).
    \end{array}
\end{equation}
 Here $a = \frac{{{N^2}{\theta ^2}}}{{2}}a'$ and $b = \frac{{{N^2}{\theta ^2}}}{{2}}b'$. Now, by comparing
 Eqs.(~\ref{equation.ori}) and (~\ref{equation.crosssection}), we can determine gluon contributions to the
 nucleon structure function which are denoted by $w_1^{gluon}$ and $w_2^{gluon}$ respectively:
\begin{equation}\label{equation.w1gluon}
    w_1^{gluon} = \frac{b}{2}\delta (\nu  - \frac{{{Q^2}}}{{2M\,{x_g}}}),
\end{equation}
\begin{equation}\label{equation.w2gluon}
    w_2^{gluon} = a\,\delta (\nu  - \frac{{{Q^2}}}{{2M\,{x_g}}}),
\end{equation}
 in which we use ${m_{eff}} = {x_g}M$. Here ${x_g}$ is fraction of nucleon momentum which is carried by gluon. To
 obtain nucleon structure function which is resulted from electron-gluon scattering, it is needed to  multiply
 $w_1^{gluon}$ and $w_2^{gluon}$  by $g({x_g})$, as probability function to find gluon with fraction of nucleon's
 momentum. Then taking integrate with respect to ${x_g}$ would be resulted to:
\begin{equation}\label{equaation.integral1}
   \begin{array}{l}
{W_1}({Q^2},\nu ) = \int\limits_0^1 {d{x_g}g({x_g})w_1^{gluon}} \\
\qquad\qquad\;\; = \int\limits_0^1 {d{x_g}g({x_g})\frac{{b({x_g})}}{2}\delta (\nu  - \frac{{{Q^2}}}{{2M{x_g}}})}
\\
\qquad\qquad {\kern 1pt}  = \int\limits_0^1 {d{x_g}g({x_g})\frac{{b({x_g})}}{2}\frac{{{x_g}}}{\nu }\delta ({x_g}
- x)} \\
\qquad\qquad{\kern 1pt}  = \frac{{b(x)}}{2}\frac{x}{\nu }g(x) = \frac{{b(x)}}{M}\frac{{{M^2}x}}{{{Q^2}}}g(x),
\end{array}
\end{equation}
\begin{equation}\label{equation.integral2}
    \begin{array}{l}
{W_2}({Q^2},\nu ) = \int\limits_0^1 {d{x_g}g({x_g})w_2^{gluon}}\\
\qquad\qquad \;\; = \int\limits_0^1 {d{x_g}g({x_g})\,a({x_g})\,\delta (\nu  - \frac{{{Q^2}}}{{2M{x_g}}})} \\
\qquad\qquad {\kern 1pt}  = \int\limits_0^1 {d{x_g}g({x_g})\,a({x_g})\frac{{{x_g}}}{\nu }\delta ({x_g} - x)} \\
\qquad\qquad{\kern 1pt}  = \frac{1}{\nu }a(x)\,x\,g(x),
\end{array}
\end{equation}
 where Eqs.(~\ref{equation.structure function eg1},~\ref{equation.structure function eg2}) as the corrected
 portion of structure function are coming from gluon-photon interaction.

\section*{Acknowledgement}
The authors  acknowledge Yazd university to provide the required facilities to do this project. We are indebted M. Haghighat for useful discussions. We are also grateful M. M. Ettefaghi for productive
comments. We are finally thankful M.M.
Sheikh Jabbari for the  critical remarks.\\


\end{document}